\documentclass[pdflatex,sn-nature,oneside]{sn-jnl}

\usepackage{graphicx}%
\usepackage{amsmath,amssymb,amsfonts}%
\usepackage{textcomp}%

\unnumbered

\begin{document}

\title{Quantum tomography of bright squeezed vacuum via strong-field ionization}

\author[1]{\fnm{Peizeng} \sur{Li}}\equalcont{These authors contributed equally to this work.}
\author[1]{\fnm{Xiaoxiao} \sur{Long}}\equalcont{These authors contributed equally to this work.}
\author[1]{\fnm{Haodong} \sur{Liu}}
\author[1]{\fnm{Xi} \sur{Li}}
\author[1]{\fnm{Yongkai} \sur{Deng}}
\author[1]{\fnm{Chuan} \sur{Cheng}}
\author*[1,2,3]{\fnm{Yunquan} \sur{Liu}}\email{yunquan.liu@pku.edu.cn}

\affil[1]{\orgdiv{State Key Laboratory for Mesoscopic Physics and Frontiers Science Center for Nano-Optoelectronics, School of Physics}, \orgname{Peking University}, \orgaddress{\city{Beijing}, \country{China}}}
\affil[2]{\orgdiv{Collaborative Innovation Center of Extreme Optics}, \orgname{Shanxi University}, \orgaddress{\city{Taiyuan}, \state{Shanxi}, \country{China}}}
\affil[3]{\orgdiv{Jiangsu Key Laboratory of Advanced Optoelectronic Integrated Chips}, \orgname{Peking University Yangtze Delta Institute of Optoelectronics}, \orgaddress{\city{Nantong}, \country{China}}}

\abstract{Strong-field physics uses intense optical fields to probe and control electron motion on ultrafast timescales. Recent advances in bright squeezed vacuum have brought macroscopic quantum states of light into the ultrafast strong-field regime, opening prospects for high-harmonic generation and photoionization. Yet linking its phase-space structure to subcycle electron dynamics remains challenging. Here, we demonstrate strong-field tunneling ionization of xenon using a phase-referenced combination of coherent and squeezed fields, with the coherent field defining the ionization clock and quadrature fluctuations of the zero-mean squeezed field providing control. The squeezed field selectively modulates atomic tunneling barriers and electron wave packets, imprinting signatures on photoelectron momentum distributions, number statistics, and energy spectra. Using coincidence measurements of photon number and phase-sensitive photoelectron momentum distributions, we perform quantum state tomography of the driving field by reconstructing its Wigner function. These results establish a bidirectional interface between quantum optics and nonperturbative strong-field electron dynamics, in which the driving quantum field shapes subcycle electron motion while photoionization provides a phase-sensitive probe of the field's quantum state. This work opens new routes toward controlling electron dynamics with quantum light and phase-sensitive quantum state metrology through ultrafast strong-field electron dynamics.}

\maketitle
\newpage

\section*{Main}
A central goal of quantum optics is to understand how quantum states of light interact with matter and ultimately control its dynamics~\cite{dorfman-NonlinearOpticalSignals-2016,raimond-ColloquiumManipulatingQuantum-2001}. The Jaynes-Cummings model reduces this broad problem to its minimal ingredients---a two-level system coupled to a single quantized mode---yet has become a cornerstone of quantum optics~\cite{jaynes-ComparisonQuantumSemiclassical-1963,shore-JaynesCummingsModel-1993,knight-EnduringRelevanceJaynes-2024}. The model embodies a general principle: the quantum state of light can govern matter dynamics. By contrast, strong-field ionization and attosecond science have developed largely within a semiclassical framework, in which coherent-state fields are treated as deterministic waveforms for probing and controlling electron wave packets~\cite{krausz-AttosecondPhysics-2009,kienberger-SteeringAttosecondElectron-2002,han-AttoclockPhotoelectronInterferometry-2018,ge-SpatiotemporalImagingShaping-2024a}. The advent of bright nonclassical light is extending this framework by introducing the quantum state of light as an active degree of freedom in nonperturbative light-matter interactions~\cite{perez-BrightSqueezedvacuumSource-2014,lyu-GenerationSingleModeMultimode-2026a,eventzur-PhotonstatisticsForceUltrafast-2023a,gorlach-HighharmonicGenerationDriven-2023b,li-ControlTunnelingIonization-2026a}. Recent high-harmonic generation (HHG) experiments have shown that the quantum properties of bright squeezed vacuum (BSV) can be imprinted on the emitted harmonics, both when BSV serves as the primary driving field and when it acts as a quantum perturbation to a coherent-state driver~\cite{luo-GenerationBrightQuantum-2026,lemieux-PhotonBunchingHighharmonic-2025a,lyu-AttosecondQuantumSpectroscopy-2026a,rasputnyi-HighharmonicGenerationBright-2024a,tzur-AttosecondresolvedQuantumFluctuations-2025a}. Attosecond pulses carrying these BSV-derived quantum fluctuations have subsequently been employed in attosecond transient absorption spectroscopy~\cite{mor-AttosecondTransientQuantum-2026a}. Collectively, these advances establish that strong-field nonlinearities can transduce the quantum properties of a driving field into high-energy photonic observables. These developments naturally raise corresponding questions in strong-field ionization: how the quantum properties of the driving field can likewise be transferred to photoionization dynamics to shape the formation and coherent superposition of subcycle electron wave packets, and whether the resulting signatures can be extracted from electron observables.

BSV provides a particularly promising platform for exploring the role of quantum light in ultrafast strong-field physics. Previous experiments on the interaction of near-single-mode BSV with metallic nanotips~\cite{heimerl-MultiphotonElectronEmission-2024a,heimerl-QuantumLightDrives-2025e} and atoms~\cite{liu-StrongfieldIonizationAtoms-2026a,liuprl2026pgkx-8t5c} have taken an important first step by showing that its photon statistics can alter strong-field electron emission and leave measurable signatures in electron-number distributions, energy spectra, ionization yields, and momentum-space holographic structures. However, in such BSV-only configurations, photoionization is temporally confined to narrow temporal windows near the extrema of the quantum light~\cite{keldysh-IonizationFieldStrong-2023,ammosov-TunnelIonizationComplex-}. Without an independent coherent-state phase reference, the effects of different phase projections of the BSV pulses on subcycle ionization channels cannot be resolved. Moreover, whether the phase-space distribution of the driving BSV field can be reconstructed experimentally through photoionization has not been addressed. Indeed, \textit{in situ} characterization of BSV remains particularly challenging because of its high photon number and sensitivity to optical loss. These considerations raise two central questions: how quantum fluctuations of a zero-mean field can selectively modulate subcycle electron wave packets and thereby reshape their coherent superposition, and, conversely, whether phase-space fingerprints of the driving quantum field can be extracted from the signature of photoionization.

Here, we demonstrate phase-space mapping of BSV through strong-field ionization. An 800 nm linearly polarized coherent-state field launches a reproducible train of tunneling electron wave packets at well-defined subcycle times, whereas a 1600 nm BSV field polarized parallel to the coherent-state field serves as a quantum perturbation with zero mean field amplitude and produces no observable strong-field ionization on its own. Using a modified cold-target recoil ion momentum spectroscopy (COLTRIMS) apparatus~\cite{dorner-ColdTargetRecoil-2000,ullrich-RecoilionElectronMomentum-2003}, we record the photoelectron momentum, the BSV intensity $I_{\alpha}$, and the relative phase $\phi_{\mathrm{rel}}$ between the two fields on a shot-by-shot basis, allowing us to resolve strong-field ionization of Xe atoms as a joint function of relative phase and BSV intensity. By developing a Coulomb-corrected quantum-optical strong-field approximation (Q-CCSFA)~\cite{lyu-EffectPhotonQuantum-2025a,popruzhenko-CoulombcorrectedQuantumTrajectories-2008}, we show that the BSV field selectively modulates subcycle ionization saddle points and reshapes the associated tunneling electron wave packets. The signatures of this modulation are encoded into the photoelectron momentum distribution (PMD), electron-number statistics, and energy spectra. These signatures are consistently observed in both experiments and simulations. By tuning $\phi_{\mathrm{rel}}$, we control the precise timing within the optical cycle when the fluctuations of BSV act on the electron wave packets, thereby transforming an otherwise uncontrolled influence into phase-tagged subcycle modulation and enabling phase-selective control of coherent electron dynamics. Crucially, the photoelectron signature not only reveals how the driving quantum field reshapes the electron wave packets but also provides access to the field's phase-space structure. By combining BSV intensity slicing with the phase-sensitive photoelectron momentum, we reconstruct the Wigner function of the driving quantum field within the macroscopic-field approximation. We thus show that the quantum properties of the driving light can be transferred into measurable electron observables and that strong-field ionization provides a novel approach to quantum state tomography of BSV.

\begin{figure}[!htbp]
	\centering
	\includegraphics[width=0.6\linewidth]{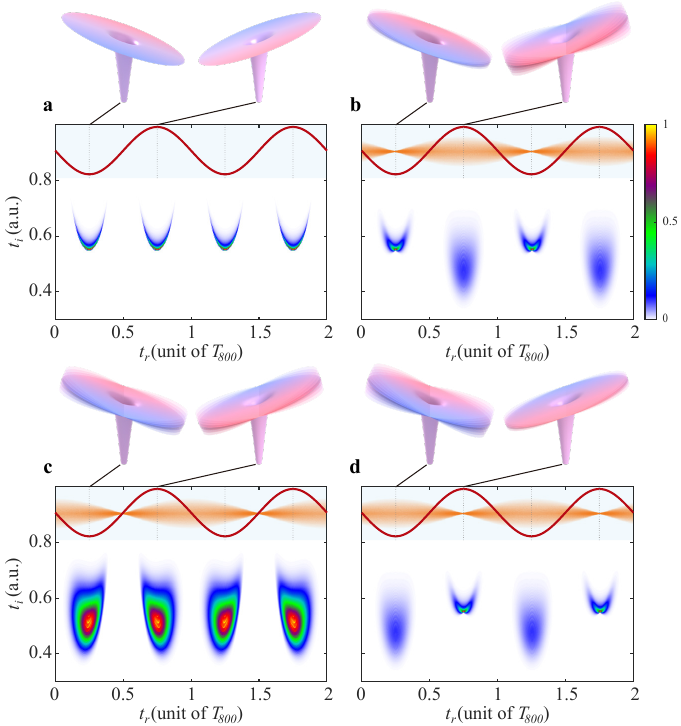}
	\caption{\textbf{Phase-dependent modulation of subcycle tunneling dynamics by BSV.}
		In each panel, the lower color map shows the distribution of complex saddle-point times $t_s=t_r+i t_i$. The red curve represents the waveform of the 800 nm coherent-state field, whereas the orange shading in panels (\textbf{b}) to (\textbf{d}) represents the distribution of the 1600 nm BSV field fluctuations. The schematics above the color maps depict the instantaneous tunneling barriers associated with the first two subcycle ionization channels. Each saddle-point distribution is normalized to its own maximum. (\textbf{a}) Saddle-point distribution for ionization driven by the 800 nm coherent-state field alone. (\textbf{b} to \textbf{d}) Corresponding saddle-point distributions in the presence of the 1600 nm BSV field for relative phases $\phi_{\mathrm{rel}}=-\pi/2$, 0, and $\pi/2$, respectively. Relative to the single-color case, the quantum fluctuations of the BSV field reshape the saddle-point distributions in a phase-dependent and subcycle-selective manner.}
	\label{fig:fig1}
\end{figure}

In Figure 1, we illustrate the time-domain picture of how a zero-mean BSV field modulates subcycle tunneling dynamics. Within the quantum-optical strong-field approximation~\cite{lyu-EffectPhotonQuantum-2025a}, each subcycle ionization channel is characterized by a complex saddle-point time $t_{s}=t_{r}+i t_{i}$, obtained by solving the saddle-point equation. Its real part $t_{r}$ specifies the electron release time, whereas its imaginary part $t_{i}$ governs the tunneling weight~\cite{popruzhenko-CoulombcorrectedQuantumTrajectories-2008,lewenstein-TheoryHighharmonicGeneration-1994}. Together with the subsequent continuum trajectory, each saddle point therefore determines the momentum and phase evolution of the corresponding electron wave packet. Figure~\ref{fig:fig1}a shows the saddle-point structure of tunneling ionization driven by the 800 nm coherent-state field alone, with four dominant ionization channels occurring over two optical cycles. When the 1600 nm BSV field is introduced, its mean electric field remains zero, but its quantum fluctuations reshape the instantaneous tunneling barrier. At a relative phase of $-\pi/2$, the first and third saddle points in Fig.~\ref{fig:fig1}b undergo similar modulation, whereas the second and fourth experience a distinct and stronger modulation. Consequently, the tunneling barriers associated with electron wave packets released in adjacent half cycles are reshaped in dramatically different ways. At a relative phase of 0, the BSV field affects the first and second subcycle ionization channels nearly identically, as shown in Fig.~\ref{fig:fig1}c. When the relative phase is shifted to $\pi/2$, Fig.~\ref{fig:fig1}d shows a complementary modulation pattern, in which the ionization channels associated with the first and third saddle points are more strongly affected. Figure~\ref{fig:fig1} thus provides a microscopic time-domain picture of how the quantum properties of the driving field reshape the tunneling barriers and associated electron wave packets, foreshadowing the observable signatures of this subcycle modulation in the final PMDs and photoelectron-number statistics.

\begin{figure}[!htbp]
	\centering
	\includegraphics[width=0.6\linewidth]{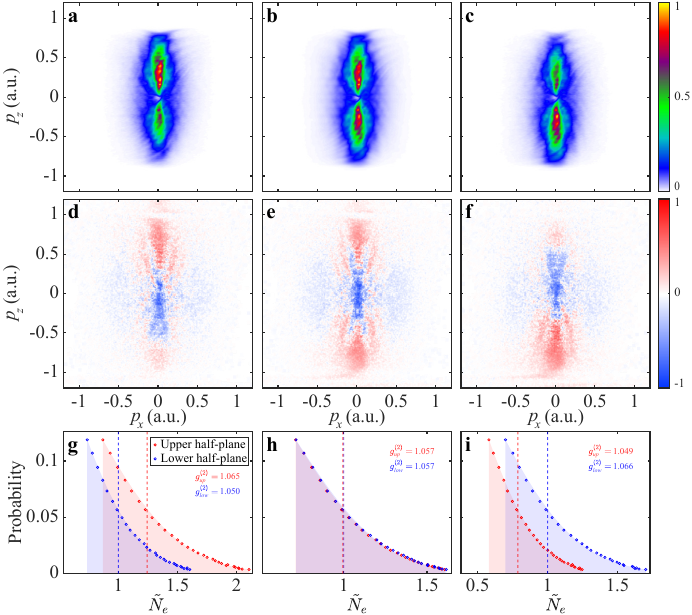}
	\caption{\textbf{Phase-selective modulation of photoelectron momentum distributions and photoelectron-number statistics by BSV.}
		The three columns correspond, from left to right, to the three representative relative phases $\phi_{1}=1.12\pi$, $\phi_{2}=0.62\pi$, and $\phi_{3}=0.12\pi$, respectively. (\textbf{a} to \textbf{c}) Experimental PMDs obtained within phase windows centered at these three representative relative phases. Each PMD is independently normalized to its own maximum. (\textbf{d} to \textbf{f}) Corresponding momentum-resolved shape-correlation maps $C_{\mathrm{shape}}(\mathbf{p})$, which quantify, at each momentum, the correlation between the normalized electron yield and shot-to-shot variations in BSV intensity. Positive (red) and negative (blue) values identify momentum regions in which the normalized electron yield is correlated and anticorrelated, respectively, with the BSV intensity fluctuations. (\textbf{g} to \textbf{i}) Probability distributions $P(\tilde{N}_{e})$ of the normalized photoelectron number in the upper ($p_z>0$, red) and lower ($p_z<0$, blue) momentum half-planes; the values of $g_{\mathrm{up}}^{(2)}$ and $g_{\mathrm{low}}^{(2)}$ are indicated. The vertical dashed lines indicate the mean normalized photoelectron number for each distribution.}
	\label{fig:fig2}
\end{figure}

Figure~\ref{fig:fig2}a--c shows the measured PMDs at three representative relative phases between the two fields, $\phi_{1}$, $\phi_{2}$, and $\phi_{3}$, highlighting the phase-dependent variation in electron emission; e.g., emission occurs predominantly in the upper half-plane at $\phi_{1}$, is nearly balanced between the two half-planes at $\phi_{2}$, and occurs predominantly in the lower half-plane at $\phi_{3}$. The phase-dependent asymmetry demonstrates that $\phi_{\mathrm{rel}}$ controls how different phase projections of the BSV field act on distinct subcycle electron wave packets, with the resulting modulation encoded in the final PMD. However, an ensemble-averaged PMD captures only the mean electron response and cannot reveal how shot-to-shot variations in BSV intensity are encoded across momentum space. We therefore evaluate a momentum-resolved shape correlation between the shot-resolved BSV intensity $I_{\alpha}$ and the normalized PMD $F_{\alpha}(\mathbf{p})$:
\begin{equation}
C_{\mathrm{shape}}(\mathbf{p})=\frac{\operatorname{Cov}\left[I_{\alpha},F_{\alpha}(\mathbf{p})\right]}{\sigma_{I_{\alpha}}\sigma_{F_{\alpha}}(\mathbf{p})}.
\label{eq:shape_correlation}
\end{equation}
\noindent
The magnitude and sign of $C_{\mathrm{shape}}$ identify momentum regions most sensitive to variations in BSV intensity and indicate whether their fractional contribution to the total electron yield increases or decreases with $I_{\alpha}$. Figure~\ref{fig:fig2}d--f shows that these correlation patterns are strongly phase selective. At $\phi_{1}$, the positive correlation is concentrated mainly in the upper half-plane, whereas at $\phi_{3}$ it shifts to the lower half-plane. At $\phi_{2}$, the correlation pattern is nearly symmetric between the two half-planes, indicating a redistribution of the normalized electron yield from the central region toward larger momenta. Notably, spider-like photoelectron holographic structures remain discernible in the correlation maps, in agreement with the study of monochromatic BSV-driven strong-field ionization~\cite{liu-StrongfieldIonizationAtoms-2026a}. This observation shows that BSV fluctuations do not simply wash out strong-field holographic interference; rather, they modulate the electron momentum distribution in a phase-selective manner while partially preserving the interference structure.

This phase-selective mapping of the BSV fluctuations is also reflected in the photoelectron number statistics. For driving by coherent-state fields with fixed component intensities and relative phase, the ionization process is characterized by a single, well-defined electron yield. When quantum-light driving is included, however, shot-to-shot intensity fluctuations of the BSV field generate a range of conditional mean electron yields, giving rise to a measurable distribution $P(\tilde{N}_{e})$ of normalized photoelectron numbers. For each BSV intensity interval within a given relative-phase window, we separately integrate the electron yields over the upper and lower halves of the corresponding PMD and construct the normalized photoelectron-number distributions $P_{\mathrm{up}}(\tilde{N}_{e})$ and $P_{\mathrm{low}}(\tilde{N}_{e})$ (see Methods for details). The corresponding second-order correlations are quantified by $g_{\mathrm{up}}^{(2)}$ and $g_{\mathrm{low}}^{(2)}$, as shown in Fig.~\ref{fig:fig2}g--i. At $\phi_{1}$, the upper-half-plane distribution exhibits a more pronounced high-yield tail, extending to larger $\tilde{N}_{e}$ than its lower-half-plane counterpart. Correspondingly, $g_{\mathrm{up}}^{(2)}$ is larger than $g_{\mathrm{low}}^{(2)}$. At $\phi_{2}$, the two distributions and their corresponding $g^{(2)}$ values are nearly identical. At $\phi_{3}$, the trend observed at $\phi_{1}$ is reversed, with the lower-half-plane distribution exhibiting a more pronounced high-yield tail and a larger $g^{(2)}$. These results show that nonlinear strong-field ionization encodes the quantum properties of the driving field not only in the shape of the PMD but also in the phase-selective photoelectron number distributions. The Q-CCSFA simulations presented in Methods reproduce these mean and fluctuation-resolved responses, supporting the interpretation that the experimental observables arise from the quantum statistical properties of the driving BSV field. Overall, these results demonstrate how the modulation of subcycle ionization saddle points reshapes the outgoing electron wave packets, producing experimentally accessible, phase-selective signatures in both the PMDs and the photoelectron-number statistics. Together, these findings confirm that the quantum properties of the driving field are imprinted on photoelectron observables.

\begin{figure}[!htbp]
	\centering
	\includegraphics[width=0.6\linewidth]{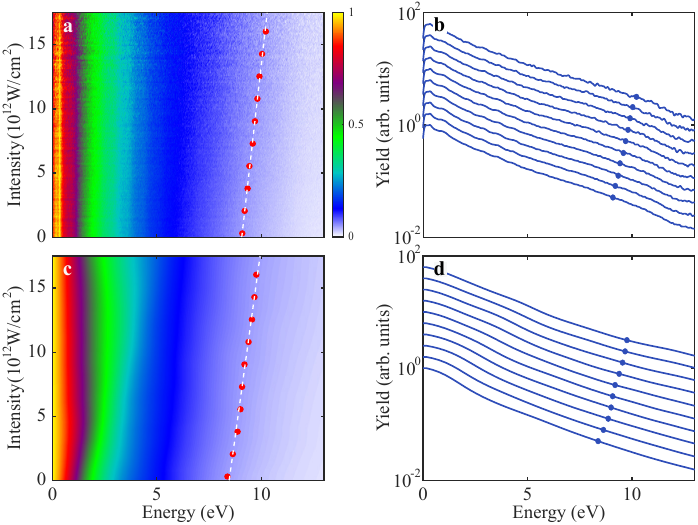}
	\caption{\textbf{Intensity-resolved modulation of photoelectron energy spectra by BSV.}
		(\textbf{a}) Experimental photoelectron energy spectra as a function of the BSV intensity $I_{\alpha}$, obtained by sorting the data into intensity slices. The intensity-dependent energy spectra are independently normalized to their respective maxima. Red circles mark the energy at which each spectrum falls to $5\%$ of its maximum. The white dashed line is a linear fit to these extracted $5\%$ energies and serves as a guide. (\textbf{b}) Representative normalized energy spectra selected from (a) and displayed with successive vertical offsets on a logarithmic yield scale to prevent overlap. The spectra are ordered from bottom to top by increasing $I_{\alpha}$, and the circles mark their corresponding $5\%$ threshold energies. (\textbf{c} and \textbf{d}) Corresponding Q-CCSFA calculations presented in the same formats as (a) and (b), respectively. Intercycle interference is omitted from the calculations to isolate the dependence of the spectral envelope and high-energy tail on $I_{\alpha}$.}
	\label{fig:fig3}
\end{figure}

Having established the mapping of BSV quantum fluctuations onto photoelectron momentum space, we then examine whether they also leave a measurable imprint on the photoelectron energy spectra. As shown in Fig.~\ref{fig:fig3}a, we define the energy at which the spectral yield falls to $5\%$ of its maximum as an operational measure of the high-energy extent of each spectrum. With increasing $I_{\alpha}$, the spectral tail extends to higher energies. The representative energy spectra plotted on a logarithmic yield scale in Fig.~\ref{fig:fig3}b make this trend more evident: those corresponding to larger values of $I_{\alpha}$ extend to higher energies. The Q-CCSFA simulations reproduce the observed intensity-dependent trend (Fig.~\ref{fig:fig3}c,d). The quantum fluctuations of the BSV field thus leave measurable imprints on the complementary photoelectron observables: momentum distributions, number statistics, and energy spectra.

\begin{figure}[!htbp]
	\centering
	\includegraphics[width=0.6\linewidth]{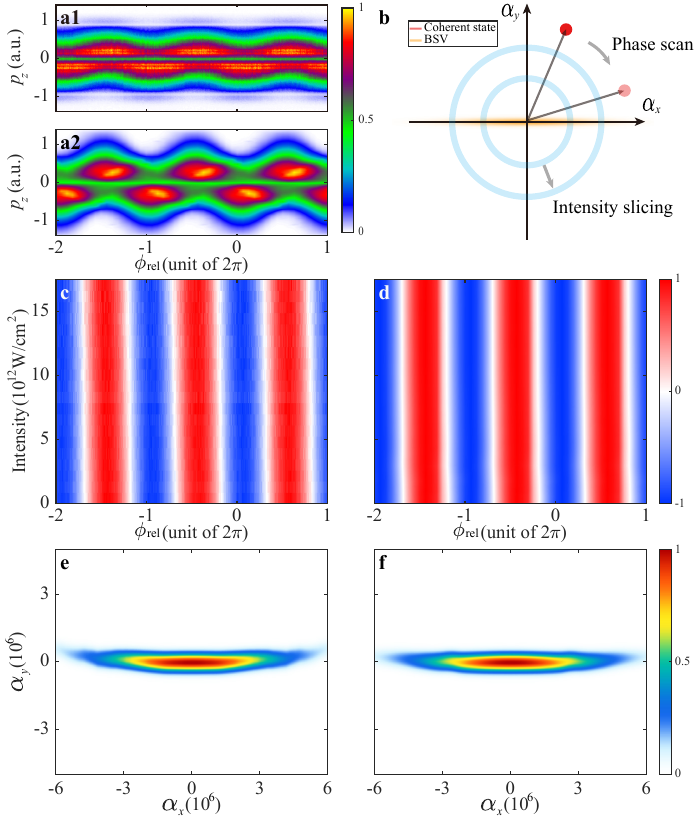}
	\caption{\textbf{Wigner-function reconstruction of BSV via phase-sensitive strong-field ionization.}
		(\textbf{a1} and \textbf{a2}) Experimental (\textbf{a1}) and Q-CCSFA (\textbf{a2}) distributions of the photoelectron momentum $p_z$ as a function of $\phi_{\mathrm{rel}}$. Both distributions are independently normalized to their respective maxima. (\textbf{b}) Schematic of the phase-space readout principle. BSV intensity slicing, followed by conversion to the corresponding field amplitude, determines the radial coordinate, whereas the phase offset extracted from the $\phi_{\mathrm{rel}}$ scan determines the angular coordinate. (\textbf{c} and \textbf{d}) Experimental (\textbf{c}) and Q-CCSFA (\textbf{d}) maps of the mean photoelectron momentum $\langle p_z\rangle$ as a function of $\phi_{\mathrm{rel}}$ and $I_{\alpha}$. (\textbf{e} and \textbf{f}) Wigner-function reconstructions of the BSV field in the $(\alpha_x,\alpha_y)$ plane obtained from the experimental data (\textbf{e}) and Q-CCSFA simulations (\textbf{f}). The radial coordinates and statistical weights are derived from the BSV intensity distribution, whereas the angular coordinates are determined by the fitted phase offsets of the corresponding $\langle p_z\rangle(\phi_{\mathrm{rel}})$ responses. Within the macroscopic-field approximation, the distribution along the squeezed quadrature approaches a delta function; for visualization, this delta-function profile is represented by a finite width equal to one tenth of the antisqueezed-quadrature width. The opposite half of the distribution is obtained by applying inversion symmetry.}
	\label{fig:fig4}
\end{figure}

Importantly, the phase-sensitive photoelectron response also provides the phase-space readout required for Wigner-function tomography of the BSV field through strong-field ionization. We continuously scan $\phi_{\mathrm{rel}}$ while measuring the photoelectron momentum $p_{z}$ along the polarization axis of the light fields. Both the experimental and simulated $p_{z}$ distributions exhibit a clear oscillatory dependence on $\phi_{\mathrm{rel}}$, as shown in Fig.~\ref{fig:fig4}a1,a2, respectively. This phase-dependent $p_{z}$ modulation provides a continuous counterpart to the PMD differences observed at the selected relative phases in Fig.~\ref{fig:fig2}. By monitoring the relative phase continuously, one can see how the quantum statistical properties of the BSV field are mapped onto distinct subcycle electron wave packets. These robust oscillations show that the ionization process maps different phase projections of the BSV field relative to the coherent-state reference field onto a measurable photoelectron momentum signal, thereby allowing us to extract the angular information about the phase space of the BSV field from the measured momentum distributions.

To translate the phase-sensitive photoelectron signatures into a phase-space representation of the driving quantum field, we associate the experimental quantities with its radial and angular coordinates in phase space. Figure~\ref{fig:fig4}b illustrates this readout principle: the shot-resolved BSV intensity is used to sort the data into intensity slices and is then converted to the corresponding field amplitude, which sets the radial coordinate, whereas the phase offset extracted from the relative-phase scan sets the angular coordinate. Using this mapping, we construct a two-dimensional map of $\langle p_{z}\rangle$ as a function of $I_{\alpha}$ and $\phi_{\mathrm{rel}}$. Figure~\ref{fig:fig4}c shows the experimental result, and Fig.~\ref{fig:fig4}d shows the corresponding Q-CCSFA result. For each BSV intensity slice, we then fit the oscillation of $\langle p_{z}\rangle$ as a function of relative phase to a sinusoidal function and extract the corresponding phase offset. Using the lowest-intensity slice as the phase reference, the differences between the fitted phase offsets determine the relative angular coordinates associated with the other BSV intensity slices. By combining the radial coordinates and statistical weights derived from the measured BSV intensity distribution with the extracted angular coordinates, we obtain a phase-space image of the BSV field in the $(\alpha_x,\alpha_y)$ plane, as shown in Fig.~\ref{fig:fig4}e. This image provides a reconstruction of the Husimi $Q$ distribution of the BSV field using the phase-sensitive photoelectron readout. The Husimi $Q$ distribution is related to the Wigner function by convolution with a Gaussian kernel of vacuum width~\cite{cahill-DensityOperatorsQuasiprobability-1969}. In the macroscopic-field limit, the BSV Husimi $Q$ distribution and the Wigner function approach the same limiting form: a delta function along the squeezed quadrature and a distribution of finite width along the antisqueezed quadrature~\cite{gorlach-HighharmonicGenerationDriven-2023b}. Within this approximation, the reconstructed Husimi $Q$ distribution provides an operationally equivalent representation of the corresponding Wigner function, thereby enabling Wigner-function tomography of the bright driving quantum field. Applying the same reconstruction procedure to the simulated photoelectron signal yields the phase-space image shown in Fig.~\ref{fig:fig4}f, which is in overall agreement with the experimental result. These results show that strong-field ionization not only encodes the quantum statistical properties of the driving quantum field in the outgoing photoelectron wave packets and associated observables but also enables the phase-space structure of the driving field to be extracted from the photoelectron response. Building on this framework, incorporating photoelectron momentum variances, covariances, and higher-order moments as additional readout observables could, in principle, provide access to higher-order fluctuations and correlations of the driving field~\cite{zhou-AttosecondAccessQuantum-2026a,stammer-AttosecondMetrologyBright-2026a}, opening a route to characterizing quantum states of light with more complex nonclassical phase-space structures.

\section*{Conclusion}

In summary, we experimentally demonstrate \textit{in situ} quantum state tomography of BSV using phase-referenced strong-field ionization. BSV serves as a controllable quantum degree of freedom for shaping outgoing photoelectron wave packets on a subcycle timescale. Although the ensemble-averaged electric field of the 1600 nm BSV vanishes, its quadrature fluctuations shift the ionization saddle points within the tunneling windows defined by the 800 nm coherent-state reference field, thereby reshaping the photoelectron wavefunction. The experimental results show that nonlinear strong-field ionization transduces the quantum statistical properties of the BSV field into measurable signatures in three complementary photoelectron observables: momentum distributions, number statistics, and energy spectra. By combining continuous relative-phase scanning with shot-resolved BSV intensity slicing, we reconstruct the Wigner function of the BSV field within the macroscopic-field approximation, using the phase-sensitive photoelectron momentum as the tomographic readout. Together, these results establish a bidirectional experimental interface between quantum optics and nonperturbative strong-field electron dynamics: the quantum fluctuations of the driving field modulate the electron dynamics, while photoionization provides a phase-sensitive metrological probe of that field. Our work opens a new route toward phase-sensitive quantum-light metrology based on subcycle electron dynamics.

\clearpage
\section{Methods}\label{sec:methods}

\renewcommand{\figurename}{Extended Data Fig.}
\renewcommand{\thefigure}{\arabic{figure}}
\renewcommand{\theHfigure}{ED.\arabic{figure}}
\setcounter{figure}{0}

\subsection{Experimental setup and characterization of the input BSV field}\label{sec:experimental-setup}

\begin{figure}[!htbp]
	\centering
	\includegraphics[width=0.9\linewidth]{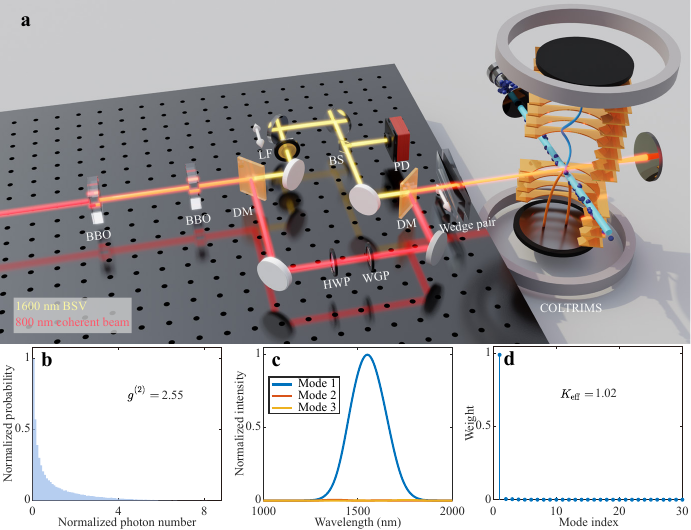}
	\caption{\textbf{Experimental setup for strong-field ionization with phase-referenced coherent-state and BSV fields.} 
		(\textbf{a}) Schematic of the experimental setup. (\textbf{b}) Photon-number distribution of the input BSV field. The horizontal axis shows the photon number normalized to its mean, $n/\langle n\rangle$, whereas the vertical axis shows the probability density normalized to its maximum. The second-order intensity correlation calculated from this distribution is $g^{(2)}=2.55$. (\textbf{c}) Spectral intensity profiles of the first three Schmidt modes. (\textbf{d}) Weights of the spectral Schmidt modes. The first Schmidt mode accounts for 0.99 of the total weight, and the full set of Schmidt weights yields an effective mode number of $K_{\mathrm{eff}}=1.02$.}
	\label{fig:s1}
\end{figure}

Extended Data Fig.~\ref{fig:s1}a shows a schematic of the experimental setup used to investigate strong-field ionization of Xe atoms driven by a hybrid field comprising a coherent-state field and a BSV field. The coherent-state field at 800 nm pumps two beta-barium borate (BBO) crystals to generate a bright squeezed vacuum (BSV) field at 1600 nm via type-I high-gain degenerate spontaneous parametric down-conversion~\cite{perez-BrightSqueezedvacuumSource-2014,lyu-GenerationSingleModeMultimode-2026a}. A dichroic mirror (DM) separates the 800 nm and 1600 nm fields into two arms. A motorized translation stage adjusts the optical path length in one arm to establish temporal overlap between the two fields in the interaction region. In the BSV arm, a beam splitter (BS) directs a small fraction of the field to a photodetector (PD), whose signal is used to determine the BSV intensity for each laser shot. The two fields are subsequently recombined at a second DM, and a wedge pair is used to tune their relative phase $\phi_{\mathrm{rel}}$. The driving light fields are focused into a cold-target recoil ion momentum spectroscopy (COLTRIMS) apparatus, where they interact with Xe atoms. In the interaction region, the two fields are linearly polarized along a common direction. We define the common polarization direction as the $z$ axis, the laser propagation direction as the $y$ axis, and the gas-jet direction, which is perpendicular to both, as the $x$ axis. For a BSV field realization labeled by $\alpha$, the total driving field can be written as
\begin{equation}
\mathbf{E}_{\alpha}(t)=\left[E_{800}\sin\left(\omega_{800}t+\phi_{\mathrm{rel}}\right)+E_{1600,\alpha}\sin\left(\omega_{1600}t\right)\right]\hat{\mathbf{z}}.
\label{eq:driving_field}
\end{equation}
\noindent
Here, $E_{800}$ and $E_{1600,\alpha}$ are the corresponding field amplitudes, $\omega_{800}$ and $\omega_{1600}$ are their angular frequencies, and $\phi_{\mathrm{rel}}$ is the relative phase between the two fields. In the experiment, the intensity of the 800 nm coherent-state field is $I_{800}=5.7\times10^{13}~\mathrm{W/cm^{2}}$, and the mean intensity of the 1600 nm BSV field is $I_{1600}=3.5\times10^{12}~\mathrm{W/cm^{2}}$. Shot-resolved measurements of the BSV intensity and relative phase are synchronized with the corresponding photoelectron momentum measurements, providing the experimental basis for the phase-resolved PMDs, BSV intensity slicing, shape-correlation analysis, photoelectron-number statistics, energy spectra, and driving-field phase-space readout presented in the main text.

We next characterize the photon-number statistics and spectral mode structure of the input BSV field. The photon-number distribution inferred from the shot-resolved BSV intensity measurements is strongly asymmetric and exhibits a pronounced long tail [Extended Data Fig.~\ref{fig:s1}b], reflecting the large shot-to-shot intensity fluctuations of the BSV field. The second-order intensity correlation calculated from this distribution is $g^{(2)}=2.55$, which indicates strong photon bunching and is close to the theoretical high-photon-number limit of $g^{(2)}=3$ for single-mode BSV~\cite{sh.iskhakov-SuperbunchedBrightSqueezed-2012a}. To characterize its spectral mode content, we perform a Schmidt decomposition of the measured BSV spectra~\cite{barakat-SimultaneousMeasurementMultimode-2025} and extract the spectral Schmidt modes and their respective weights [Extended Data Fig.~\ref{fig:s1}c,d]. The first Schmidt mode accounts for 0.99 of the total weight, and the complete set of Schmidt weights yields an effective mode number of $K_{\mathrm{eff}}=1.02$, indicating only minor contributions from higher-order spectral modes. Taken together, the photon-number statistics and spectral mode decomposition support the characterization of the experimentally generated 1600 nm field as an effectively single-mode BSV field.

\subsection{Numerical simulations of phase-selective photoelectron modulation}\label{sec:numerical-simulations}

\begin{figure}[!htbp]
	\centering
	\includegraphics[width=0.6\linewidth]{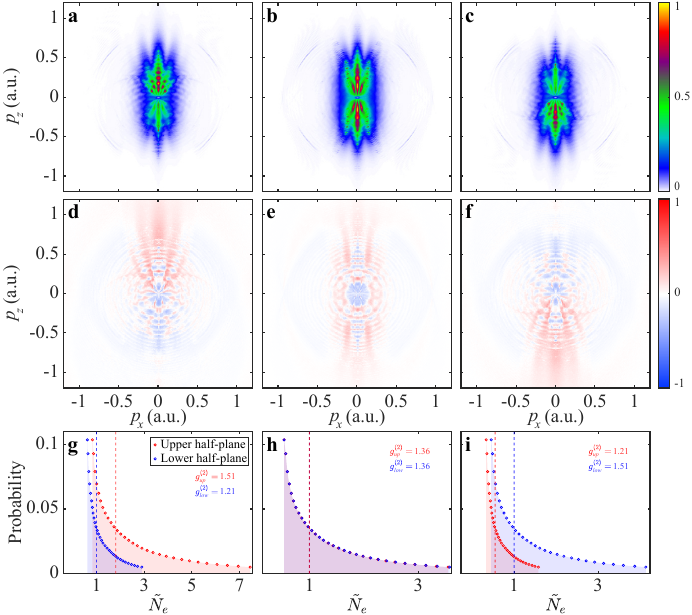}
	\caption{\textbf{Q-CCSFA simulations of phase-selective photoelectron modulation.}
		From left to right, the columns correspond to the same three representative relative phases between the two fields, $\phi_{1}$, $\phi_{2}$, and $\phi_{3}$, as in Fig.~\ref{fig:fig2}. (\textbf{a} to \textbf{c}) Simulated PMDs at these phases. Each PMD is independently normalized to its own maximum. (\textbf{d} to \textbf{f}) Corresponding momentum-resolved shape-correlation maps $C_{\mathrm{shape}}(\mathbf{p})$. (\textbf{g} to \textbf{i}) Normalized upper- and lower-half-plane photoelectron-number distributions $P_{\mathrm{up}}(\tilde{N}_{e})$ and $P_{\mathrm{low}}(\tilde{N}_{e})$ for the same three relative phases. The vertical dashed lines indicate the mean normalized photoelectron number for each distribution, and the corresponding second-order correlation values $g_{\mathrm{up}}^{(2)}$ and $g_{\mathrm{low}}^{(2)}$ are indicated in each panel.}
	\label{fig:s2}
\end{figure}

Extended Data Fig.~\ref{fig:s2} presents Q-CCSFA simulations that reproduce the phase-selective photoelectron modulation shown in Fig.~\ref{fig:fig2}. The simulations are performed at the same three representative relative phases between the two fields, $\phi_{1}$, $\phi_{2}$, and $\phi_{3}$, as in the experiment. Extended Data Fig.~\ref{fig:s2}a--c shows the simulated PMDs at these phases, with electron emission concentrated predominantly in the upper half-plane at $\phi_{1}$, nearly balanced between the two half-planes at $\phi_{2}$, and concentrated predominantly in the lower half-plane at $\phi_{3}$. Extended Data Fig.~\ref{fig:s2}d--f shows the corresponding momentum-resolved shape-correlation maps, and Extended Data Fig.~\ref{fig:s2}g--i shows the normalized upper- and lower-half-plane photoelectron-number distributions $P_{\mathrm{up}}(\tilde{N}_{e})$ and $P_{\mathrm{low}}(\tilde{N}_{e})$ at the corresponding relative phases.

Specifically, the maps in Extended Data Fig.~\ref{fig:s2}d--f quantify how shot-to-shot variations in BSV intensity reshape the PMD. For each relative phase, we obtain a photoelectron momentum distribution $Y_{\alpha}(\mathbf{p})$ for each BSV intensity interval indexed by $\alpha$. To isolate changes in the PMD shape from variations in the total ionization yield, we normalize each distribution to unit total yield:
\begin{equation}
F_{\alpha}(\mathbf{p})=
\frac{Y_{\alpha}(\mathbf{p})}
{\sum_{\mathbf{p}^{\prime}}Y_{\alpha}(\mathbf{p}^{\prime})}.
\label{eq:yield}
\end{equation}
\noindent
Using the statistical weight and effective BSV intensity $I_{\alpha}$ of each interval, we then evaluate the bin-weighted, momentum-resolved shape correlation between the BSV intensity and the normalized PMD:
\begin{equation}
C_{\mathrm{shape}}(\mathbf{p})=
\frac{\operatorname{Cov}\left[I_{\alpha},F_{\alpha}(\mathbf{p})\right]}
{\sigma_{I_{\alpha}}\sigma_{F_{\alpha}}(\mathbf{p})}.
\label{eq:sup_shape_correlation}
\end{equation}
\noindent
Here, $\operatorname{Cov}$ and $\sigma$ denote the weighted covariance and the weighted standard deviation, respectively. Thus, $C_{\mathrm{shape}}(\mathbf{p})$ characterizes how the BSV intensity fluctuations redistribute the normalized electron yield across momentum space. A positive correlation indicates that the fraction of the total electron yield contributed by a given momentum region increases with the BSV intensity, whereas a negative correlation indicates that this fraction decreases.

We next construct the photoelectron number statistics shown in Extended Data Fig.~\ref{fig:s2}g--i. For each relative phase and BSV intensity interval indexed by $\alpha$, we separately integrate the electron yield over the upper and lower halves of the corresponding PMD to obtain the conditional mean electron yields $\mu_{\alpha}^{\mathrm{up}}$ and $\mu_{\alpha}^{\mathrm{low}}$. We map these yields onto photoelectron numbers using a scale factor $A$, such that $N_{e,\alpha}^{\mathrm{up/low}}=A\mu_{\alpha}^{\mathrm{up/low}}$. We define the normalized photoelectron number plotted along the horizontal axis as
\begin{equation}
\tilde{N}_{e}^{\mathrm{up/low}}
=
\frac{N_{e,\alpha}^{\mathrm{up/low}}}
{\left\langle N_{e}^{\mathrm{low}}\right\rangle},
\label{eq:normalized_electron_number}
\end{equation}
\noindent
where $\left\langle N_{e}^{\mathrm{low}}\right\rangle$ denotes the mean lower-half-plane photoelectron number evaluated separately for each relative phase. Because the same scale factor $A$ appears in both the numerator and denominator, it cancels from the normalized quantity. The resulting distributions reproduce the phase-dependent behavior observed experimentally. At $\phi_{1}$, the upper-half-plane distribution exhibits a more pronounced high-yield tail, extending to larger values of $\tilde{N}_{e}$ than its lower-half-plane counterpart. At $\phi_{2}$, the two distributions are nearly identical, whereas at $\phi_{3}$, the behavior observed at $\phi_{1}$ is reversed.

When analyzed using the same procedure as that applied to the experimental data, the Q-CCSFA results reproduce the phase-selective PMD asymmetry, the momentum-resolved shape-correlation maps, and the phase-dependent differences between the upper- and lower-half-plane photoelectron-number distributions shown in Fig.~\ref{fig:fig2}. The agreement across these observables shows that BSV quantum fluctuations do not merely alter the total ionization yield. Furthermore, $\phi_{\mathrm{rel}}$ controls how these fluctuations are projected onto different subcycle electron wave packets, giving rise to corresponding mean and fluctuation-resolved signatures in the PMDs and photoelectron number statistics.

\subsection{Continuous relative-phase control of subcycle photoelectron wave-packet dynamics}\label{sec:phase-control}

\begin{figure}[!htbp]
	\centering
	\includegraphics[width=0.6\linewidth]{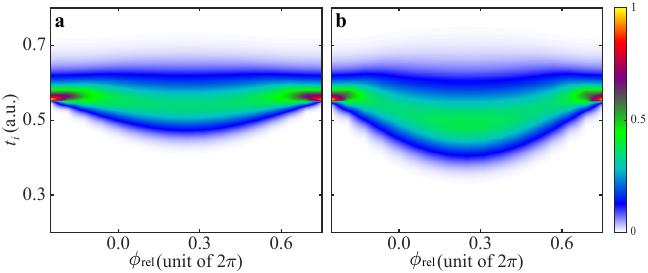}
	\caption{\textbf{Subcycle photoelectron wave-packet dynamics as a function of the relative phase between the coherent-state and BSV fields.}
		The color maps show the distributions of $t_i$, the imaginary part of the complex saddle-point time associated with the formation of the first subcycle photoelectron wave packet, as functions of the relative phase $\phi_{\mathrm{rel}}$ between the coherent-state and BSV fields. Each distribution is independently normalized to its own maximum. (\textbf{a} and \textbf{b}) Results obtained at mean BSV intensities of $I_{1600}=1.3\times10^{12}~\mathrm{W/cm^{2}}$ and $3.5\times10^{12}~\mathrm{W/cm^{2}}$, respectively. All other simulation parameters are held fixed. Increasing the mean BSV intensity enhances the amplitude of the phase-dependent modulation of the $t_i$ distribution.}
	\label{fig:s3}
\end{figure}

Extended Data Fig.~\ref{fig:s3} highlights the dual role of the relative phase $\phi_{\mathrm{rel}}$ in strong-field ionization driven by a hybrid field comprising a coherent-state field and a BSV field. In Fig.~\ref{fig:fig4}, $\phi_{\mathrm{rel}}$ serves as a scan coordinate for reading out phase-space information about the driving field from the photoelectron momentum distribution; in Extended Data Fig.~\ref{fig:s3}, it acts as a control parameter that continuously modulates the formation dynamics of a subcycle photoelectron wave packet. Specifically, in the simulations, we continuously scan $\phi_{\mathrm{rel}}$ and extract the distribution of the imaginary component $t_i$ of the complex saddle-point time associated with the formation of the first subcycle photoelectron wave packet. Panels a and b of Extended Data Fig.~\ref{fig:s3} correspond to two different mean BSV intensities, with all other simulation parameters held fixed. At each intensity, scanning $\phi_{\mathrm{rel}}$ systematically modifies the $t_i$ distribution of the first subcycle wave packet, demonstrating continuous relative-phase control over its formation dynamics. A comparison between panels a and b further shows that increasing the mean BSV intensity enhances the amplitude of the phase-dependent modulation, consistent with the increased magnitude of the BSV field fluctuations producing a stronger perturbation of the subcycle wave-packet dynamics. Together, Fig.~\ref{fig:fig4} and Extended Data Fig.~\ref{fig:s3} establish $\phi_{\mathrm{rel}}$ as both a phase-space readout coordinate for the driving quantum field and a control parameter for subcycle photoelectron wave-packet dynamics.

\clearpage

\bibliography{Phase_referenced_quantum_light}

@article{cahill-DensityOperatorsQuasiprobability-1969,
  author  = {Cahill, K. E. and Glauber, R. J.},
  title   = {Density operators and quasiprobability distributions},
  journal = {Phys. Rev.},
  volume  = {177},
  pages   = {1882--1902},
  year    = {1969},
  doi     = {10.1103/PhysRev.177.1882}
}

@article{liuprl2026pgkx-8t5c,
  author  = {Liu, Haodong and Long, Xiaoxiao and Li, Peizeng and Liu, Yunquan},
  title   = {Energy- and momentum-resolved single-shot yield fluctuations in bright-squeezed-vacuum-driven atomic strong-field ionization},
  journal = {Phys. Rev. Lett.},
  year    = {2026},
  doi     = {10.1103/pgkx-8t5c},
  url     = {https://doi.org/10.1103/pgkx-8t5c}
}

@article{li-ControlTunnelingIonization-2026a,
  author  = {Li, Peizeng and Long, Xiaoxiao and Lyu, Zijian and Liu, Haodong and Dou, Yankun and Qin, Ziheng and Ge, Peipei and Liu, Yunquan},
  title   = {Control of tunneling ionization and subcycle interference dynamics with squeezed coherent light},
  journal = {Phys. Rev. Res.},
  volume  = {8},
  pages   = {023047},
  year    = {2026},
  doi     = {10.1103/pbc2-3r6m}
}

@article{dorfman-NonlinearOpticalSignals-2016,
  author  = {Dorfman, Konstantin E. and Schlawin, Frank and Mukamel, Shaul},
  title   = {Nonlinear optical signals and spectroscopy with quantum light},
  journal = {Rev. Mod. Phys.},
  volume  = {88},
  pages   = {045008},
  year    = {2016},
  doi     = {10.1103/RevModPhys.88.045008}
}

@article{jaynes-ComparisonQuantumSemiclassical-1963,
  author  = {Jaynes, E. T. and Cummings, F. W.},
  title   = {Comparison of quantum and semiclassical radiation theories with application to the beam maser},
  journal = {Proc. IEEE},
  volume  = {51},
  pages   = {89--109},
  year    = {1963},
  doi     = {10.1109/PROC.1963.1664}
}

@article{knight-EnduringRelevanceJaynes-2024,
  author  = {Knight, Peter L. and Gerry, Christopher C. and Birrittella, Richard J. and Alsing, Paul M.},
  title   = {Enduring relevance of the {Jaynes--Cummings} model: a personal perspective [Invited]},
  journal = {J. Opt. Soc. Am. B},
  volume  = {41},
  pages   = {C91--C96},
  year    = {2024},
  doi     = {10.1364/JOSAB.524015}
}

@article{raimond-ColloquiumManipulatingQuantum-2001,
  author  = {Raimond, J. M. and Brune, M. and Haroche, S.},
  title   = {Manipulating quantum entanglement with atoms and photons in a cavity},
  journal = {Rev. Mod. Phys.},
  volume  = {73},
  pages   = {565--582},
  year    = {2001},
  doi     = {10.1103/RevModPhys.73.565}
}

@article{shore-JaynesCummingsModel-1993,
  author  = {Shore, Bruce W. and Knight, Peter L.},
  title   = {The {Jaynes--Cummings} model},
  journal = {J. Mod. Opt.},
  volume  = {40},
  pages   = {1195--1238},
  year    = {1993},
  doi     = {10.1080/09500349314551321}
}

@article{ge-SpatiotemporalImagingShaping-2024a,
  author  = {Ge, Peipei and Dou, Yankun and Han, Meng and Fang, Yiqi and Deng, Yongkai and Wu, Chengyin and Gong, Qihuang and Liu, Yunquan},
  title   = {Spatiotemporal imaging and shaping of electron wave functions using novel attoclock interferometry},
  journal = {Nat. Commun.},
  volume  = {15},
  pages   = {497},
  year    = {2024},
  doi     = {10.1038/s41467-024-44775-5}
}

@article{han-AttoclockPhotoelectronInterferometry-2018,
  author  = {Han, Meng and Ge, Peipei and Shao, Yun and Gong, Qihuang and Liu, Yunquan},
  title   = {Attoclock photoelectron interferometry with two-color corotating circular fields to probe the phase and the amplitude of emitting wave packets},
  journal = {Phys. Rev. Lett.},
  volume  = {120},
  pages   = {073202},
  year    = {2018},
  doi     = {10.1103/PhysRevLett.120.073202}
}

@article{kienberger-SteeringAttosecondElectron-2002,
  author  = {Kienberger, R. and Hentschel, M. and Uiberacker, M. and Spielmann, Ch. and Kitzler, M. and Scrinzi, A. and Wieland, M. and Westerwalbesloh, Th. and Kleineberg, U. and Heinzmann, U. and Drescher, M. and Krausz, F.},
  title   = {Steering attosecond electron wave packets with light},
  journal = {Science},
  volume  = {297},
  pages   = {1144--1148},
  year    = {2002},
  doi     = {10.1126/science.1073866}
}

@article{krausz-AttosecondPhysics-2009,
  author  = {Krausz, Ferenc and Ivanov, Misha},
  title   = {Attosecond physics},
  journal = {Rev. Mod. Phys.},
  volume  = {81},
  pages   = {163--234},
  year    = {2009},
  doi     = {10.1103/RevModPhys.81.163}
}

@article{eventzur-PhotonstatisticsForceUltrafast-2023a,
  author  = {{Even Tzur}, Matan and Birk, Michael and Gorlach, Alexey and Kr{\"u}ger, Michael and Kaminer, Ido and Cohen, Oren},
  title   = {Photon-statistics force in ultrafast electron dynamics},
  journal = {Nat. Photonics},
  volume  = {17},
  pages   = {501--509},
  year    = {2023},
  doi     = {10.1038/s41566-023-01209-w}
}

@article{gorlach-HighharmonicGenerationDriven-2023b,
  author  = {Gorlach, Alexey and {Even Tzur}, Matan and Birk, Michael and Kr{\"u}ger, Michael and Rivera, Nicholas and Cohen, Oren and Kaminer, Ido},
  title   = {High-harmonic generation driven by quantum light},
  journal = {Nat. Phys.},
  volume  = {19},
  pages   = {1689--1696},
  year    = {2023},
  doi     = {10.1038/s41567-023-02127-y}
}

@article{lyu-GenerationSingleModeMultimode-2026a,
  author  = {Lyu, Zijian and Du, Shaowei and Li, Jingze and Liu, Haodong and Sun, Fengxiao and He, Qiongyi and Ciappina, Marcelo and Liu, Yunquan},
  title   = {Generation of single-mode and multimode bright squeezed vacuum above $10^{14}$ photons},
  journal = {Ultrafast Sci.},
  volume  = {6},
  pages   = {0175},
  year    = {2026},
  doi     = {10.34133/ultrafastscience.0175}
}

@article{perez-BrightSqueezedvacuumSource-2014,
  author  = {P{\'e}rez, A. M. and Iskhakov, T. Sh. and Sharapova, P. and Lemieux, S. and Tikhonova, O. V. and Chekhova, M. V. and Leuchs, G.},
  title   = {Bright squeezed-vacuum source with 1.1 spatial mode},
  journal = {Opt. Lett.},
  volume  = {39},
  pages   = {2403--2406},
  year    = {2014},
  doi     = {10.1364/OL.39.002403}
}

@article{lemieux-PhotonBunchingHighharmonic-2025a,
  author  = {Lemieux, Samuel and Jalil, Sohail A. and Purschke, David N. and Boroumand, Neda and Hammond, T. J. and Villeneuve, David and Naumov, Andrei and Brabec, Thomas and Vampa, Giulio},
  title   = {Photon bunching in high-harmonic emission controlled by quantum light},
  journal = {Nat. Photonics},
  volume  = {19},
  pages   = {767--771},
  year    = {2025},
  doi     = {10.1038/s41566-025-01673-6}
}

@misc{luo-GenerationBrightQuantum-2026,
  author       = {Wang, Wentao and Sun, Yaoshun and Wang, Liyuan and Hu, Lingrui and Ding, Dajun and Tang, Xiangyu and Li, Mingxuan and Yuan, Jianmin and Luo, Sizuo},
  title        = {Generation of bright quantum high-order harmonic driven by combined coherent and bright squeezed vacuum light},
  howpublished = {Preprint at Research Square \url{https://doi.org/10.21203/rs.3.rs-10205760/v1}},
  year         = {2026},
  doi          = {10.21203/rs.3.rs-10205760/v1}
}

@misc{lyu-AttosecondQuantumSpectroscopy-2026a,
  author       = {Lyu, Zijian and Sun, Fengxiao and Yi, Sili and Li, Jingze and Liu, Haodong and He, Qiongyi and Gong, Qihuang and Ivanov, Misha and Liu, Yunquan},
  title        = {Attosecond quantum spectroscopy with entangled photon pairs},
  howpublished = {Preprint at \url{https://arxiv.org/abs/2604.06707}},
  year         = {2026},
  doi          = {10.48550/arXiv.2604.06707}
}

@article{rasputnyi-HighharmonicGenerationBright-2024a,
  author  = {Rasputnyi, Andrei and Chen, Zhaopin and Birk, Michael and Cohen, Oren and Kaminer, Ido and Kr{\"u}ger, Michael and Seletskiy, Denis and Chekhova, Maria and Tani, Francesco},
  title   = {High-harmonic generation by a bright squeezed vacuum},
  journal = {Nat. Phys.},
  volume  = {20},
  pages   = {1960--1965},
  year    = {2024},
  doi     = {10.1038/s41567-024-02659-x}
}

@misc{tzur-AttosecondresolvedQuantumFluctuations-2025a,
  author       = {{Even Tzur}, Matan and Mor, Chen and Yaffe, Noa and Birk, Michael and Rasputnyi, Andrei and Kneller, Omer and Nisim, Ido and Kaminer, Ido and Chekhova, Maria and Kr{\"u}ger, Michael and Ivanov, Misha and Dudovich, Nirit and Cohen, Oren},
  title        = {Attosecond-resolved quantum fluctuations of light and matter},
  howpublished = {Preprint at \url{https://arxiv.org/abs/2511.18362}},
  year         = {2025},
  doi          = {10.48550/arXiv.2511.18362}
}

@misc{mor-AttosecondTransientQuantum-2026a,
  author       = {Mor, Chen and Yaffe, Noa and Shonfeld, Assaf and Deutsch, Keren and {Even Tzur}, Matan and Dudovich, Nirit},
  title        = {Attosecond transient quantum fluctuations},
  howpublished = {Preprint at Research Square \url{https://doi.org/10.21203/rs.3.rs-9463610/v1}},
  year         = {2026},
  doi          = {10.21203/rs.3.rs-9463610/v1}
}

@article{heimerl-MultiphotonElectronEmission-2024a,
  author  = {Heimerl, Jonas and Mikhaylov, Alexander and Meier, Stefan and H{\"o}llerer, Henrick and Kaminer, Ido and Chekhova, Maria and Hommelhoff, Peter},
  title   = {Multiphoton electron emission with non-classical light},
  journal = {Nat. Phys.},
  volume  = {20},
  pages   = {945--950},
  year    = {2024},
  doi     = {10.1038/s41567-024-02472-6}
}

@article{heimerl-QuantumLightDrives-2025e,
  author  = {Heimerl, Jonas and Rasputnyi, Andrei and P{\"o}lloth, Jonathan and Meier, Stefan and Chekhova, Maria and Hommelhoff, Peter},
  title   = {Quantum light drives electrons strongly at metal needle tips},
  journal = {Nat. Phys.},
  volume  = {21},
  pages   = {1899--1904},
  year    = {2025},
  doi     = {10.1038/s41567-025-03087-1}
}

@misc{liu-StrongfieldIonizationAtoms-2026a,
  author       = {Liu, Haodong and Long, Xiaoxiao and Li, Peizeng and Lyu, Zijian and Liu, Yunquan},
  title        = {Strong-field ionization of atoms with bright squeezed vacuum light},
  howpublished = {Preprint at \url{https://arxiv.org/abs/2604.06703}},
  year         = {2026},
  doi          = {10.48550/arXiv.2604.06703}
}

@article{barakat-SimultaneousMeasurementMultimode-2025,
  author  = {Barakat, Ismail and Kalash, Mahmoud and Scharwald, Dennis and Sharapova, Polina and Lindlein, Norbert and Chekhova, Maria},
  title   = {Simultaneous measurement of multimode squeezing through multimode phase-sensitive amplification},
  journal = {Optica Quantum},
  volume  = {3},
  pages   = {36--44},
  year    = {2025},
  doi     = {10.1364/OPTICAQ.524682}
}

@article{ammosov-TunnelIonizationComplex-,
  author  = {Ammosov, M. V. and Delone, N. B. and Krainov, V. P.},
  title   = {Tunnel ionization of complex atoms and of atomic ions in an alternating electromagnetic field},
  journal = {Sov. Phys. JETP},
  volume  = {64},
  pages   = {1191--1194},
  year    = {1986}
}

@article{keldysh-IonizationFieldStrong-2023,
  author  = {Keldysh, L. V.},
  title   = {Ionization in the field of a strong electromagnetic wave},
  journal = {Sov. Phys. JETP},
  volume  = {20},
  pages   = {1307--1314},
  year    = {1965}
}

@article{dorner-ColdTargetRecoil-2000,
  author  = {D{\"o}rner, R. and Mergel, V. and Jagutzki, O. and Spielberger, L. and Ullrich, J. and Moshammer, R. and Schmidt-B{\"o}cking, H.},
  title   = {Cold target recoil ion momentum spectroscopy: a ``momentum microscope'' to view atomic collision dynamics},
  journal = {Phys. Rep.},
  volume  = {330},
  pages   = {95--192},
  year    = {2000},
  doi     = {10.1016/S0370-1573(99)00109-X}
}

@article{ullrich-RecoilionElectronMomentum-2003,
  author  = {Ullrich, J. and Moshammer, R. and Dorn, A. and D{\"o}rner, R. and Schmidt, L. Ph. H. and Schmidt-B{\"o}cking, H.},
  title   = {Recoil-ion and electron momentum spectroscopy: reaction-microscopes},
  journal = {Rep. Prog. Phys.},
  volume  = {66},
  pages   = {1463--1545},
  year    = {2003},
  doi     = {10.1088/0034-4885/66/9/203}
}

@article{lyu-EffectPhotonQuantum-2025a,
  author  = {Lyu, Zijian and Sun, Fengxiao and Fang, Yiqi and He, Qiongyi and Liu, Yunquan},
  title   = {Effect of photon quantum statistics on electrons in above-threshold ionization},
  journal = {Phys. Rev. Res.},
  volume  = {7},
  pages   = {L012072},
  year    = {2025},
  doi     = {10.1103/PhysRevResearch.7.L012072}
}

@article{popruzhenko-CoulombcorrectedQuantumTrajectories-2008,
  author  = {Popruzhenko, S. V. and Paulus, G. G. and Bauer, D.},
  title   = {Coulomb-corrected quantum trajectories in strong-field ionization},
  journal = {Phys. Rev. A},
  volume  = {77},
  pages   = {053409},
  year    = {2008},
  doi     = {10.1103/PhysRevA.77.053409}
}

@article{lewenstein-TheoryHighharmonicGeneration-1994,
  author  = {Lewenstein, M. and Balcou, Ph. and Ivanov, M. Yu. and L'Huillier, A. and Corkum, P. B.},
  title   = {Theory of high-harmonic generation by low-frequency laser fields},
  journal = {Phys. Rev. A},
  volume  = {49},
  pages   = {2117--2132},
  year    = {1994},
  doi     = {10.1103/PhysRevA.49.2117}
}

@article{sh.iskhakov-SuperbunchedBrightSqueezed-2012a,
  author  = {Iskhakov, T. Sh. and P{\'e}rez, A. M. and Spasibko, K. Yu. and Chekhova, M. V. and Leuchs, G.},
  title   = {Superbunched bright squeezed vacuum state},
  journal = {Opt. Lett.},
  volume  = {37},
  pages   = {1919--1921},
  year    = {2012},
  doi     = {10.1364/OL.37.001919}
}

@misc{stammer-AttosecondMetrologyBright-2026a,
  author       = {Stammer, Philipp and Rivera-Dean, Jordi and Pisanty, Emilio and Lewenstein, Maciej},
  title        = {Attosecond metrology of bright quantum light},
  howpublished = {Preprint at \url{https://arxiv.org/abs/2607.06395}},
  year         = {2026},
  doi          = {10.48550/arXiv.2607.06395}
}

@misc{zhou-AttosecondAccessQuantum-2026a,
  author       = {Zhou, En-Rui and Mao, Yi-Jia and He, Pei-Lun and He, Feng},
  title        = {Attosecond access to the quantum noise of light},
  howpublished = {Preprint at \url{https://arxiv.org/abs/2604.13485}},
  year         = {2026},
  doi          = {10.48550/arXiv.2604.13485}
}

\section*{Acknowledgements}
This project is supported by the National Key R$\&$D Program (No. 2022YFA1604301), the Fundamental and Interdisciplinary Disciplines Breakthrough Plan of the Ministry of Education of China (No. JYB2025XDXM201), the Natural Science Foundation of China (No. 12334013) and Quantum Science and Technology-National Science and Technology Major Project (Grant No. 2025ZD0301000).

\section*{Author contributions}
Y.~Liu conceived the project. P.~Li and X.~Long performed the experiments and analyzed the data. P.~Li performed the theoretical calculations. All authors discussed the results. P.~Li and Y.~Liu wrote the manuscript with input from all authors.

\section*{Competing interests}
The authors declare no competing interests.

\section*{Data availability}
All data needed to evaluate the conclusions in the paper are present in the paper and/or the Methods. Additional data are available from the corresponding authors upon request.

\section*{Code availability}
The analysis code is available from the corresponding authors upon request.

\section*{Additional information}
Correspondence and requests for materials should be addressed to Y.~Liu (yunquan.liu@pku.edu.cn).

\end{document}